\documentclass[12pt, preprint]{aastex} 
\shorttitle{Mixed Pairs} \shortauthors{Domingue et al.} 
\begin{document} 
\title{Mixed-Morphology Pairs as a Breeding Ground for Active Nuclei} 
\author{Donovan L. Domingue} 
\affil{Department of Chemistry and Physics, Georgia College and State University, CBX 082, Milledgeville, GA 31061} 
\email{donovan.domingue@gcsu.edu} 
\and
\author{Jack W. Sulentic and Adriana Durbala} 
\affil{Department of Physics and Astronomy, University of Alabama, Box 870324, Tuscaloosa, AL 35487 } 
\email{giacomo@merlot.astr.ua.edu, adriana.durbala@ua.edu} 
\begin{abstract} 
Mixed morphology pairs offer a simplification of the interaction 
equation that involves a gas-rich fast rotator paired with a gas-poor 
slow rotator. In past low resolution IRAS studies it was assumed that 
the bulk of the far infrared emission originated in the spiral 
component. However our ISO studies revealed a surprising number of 
early-type components with significant IR emission some of which turned 
out to show active nuclei. This motivated us to look at the current 
statistics of active nuclei in mixed pairs using the FIR-radio 
continuum correlation as a diagnostic. We find a clear excess of 
early-type components with radio continuum emission and an active 
nucleus. We suggest that they arise more often in mixed pairs via 
cross fueling of gas from the spiral companion. This fuel is more 
efficiently channeled into the nucleus of the slow rotating receptor. 
In a sample of 112 mixed-morphology pairs from the Karachentsev catalog 
we find that about 25-30\% of detected mixed pairs show a displacement 
from the radio-FIR relation defined by normal star forming galaxies. 
The latter objects show excess radio continuum emission while others 
extend the relation to unusually high radio and FIR flux levels. Many 
of the outliers/extreme emitters involve an early-type component with 
an active nucleus. The paired E/S0 galaxies in the sample exhibit a 
significant excess detection fraction and a marginal excess luminosity 
distribution compared to isolated unpaired E/S0 galaxies. 
\end{abstract} 
\section{Introduction} 
Binary galaxies play an important role in the study of galaxy evolution 
because they represent the most common interaction scenario and comprise 
about 10\% of galaxies in the non-clustered Universe (Xu \& Sulentic 1991). 
Gravitational interactions in pairs affect galaxy evolution because these 
encounters can: 1) redistribute gas within each galaxy, 2) enhance the level 
of star formation (Keel et al. 1985; Xu \& Sulentic 1991) and possibly 3) 
stimulate active nuclei. Between 15--25\% of a reasonably unbiased sample of 
binary galaxies involves systems with mixed (E/S0+S) morphology (e.g. 
Karachentsev 1972; Reduzzi \& Rampazzo 1995). They are perhaps the most 
puzzling and, at the same time, a useful simplification of the galaxy-galaxy 
interaction equation involving a gas-rich rapid-rotator paired with a gas-poor 
companion with low specific angular momentum. Mixed pairs minimize the role 
of the relative orientation of pair component spin vectors in driving 
interaction-induced effects (e.g. Keel 1991). The late-type spiral component 
should be the primary source of gas in a mixed pair and this is reflected 
in HI studies (see e.g. Zasov \& Sulentic 1994). It is therefore expected 
to be the site of all or most star formation and nuclear activity. The activity 
might be stimulated by gravitational perturbations from the early-type 
companion. Evidence for significant amounts of gas and/or star formation/active 
nuclei in early-type components have not been reported and are not expected.

Mid and far infrared (MIR/FIR) emission has proven to be one of the 
most sensitive diagnostics of interaction induced enhancements in pairs 
(Kennicutt et al. 1987; Sulentic 1989; Xu \& Sulentic 1991; 
Hernandez-Toledo et al. 1999,2001). It is especially useful for 
assessing the level of star formation enhancement in such pairs. Radio 
continuum observations also reveals excess detections and higher flux 
levels in binary galaxies (Sulentic 1976; Condon et al. 1982). 
Throughout the history of IR studies, largely exploiting the IRAS 
database, the working hypothesis was that MIR/FIR emission originated 
largely from late-type galaxies in unresolved pairs and groups. This 
gives mixed pairs a possible advantage over spiral-spiral pairs where 
both components are expected to produce similar levels of radio and IR 
emission. In mixed pairs we can study interaction properties and infer 
which component is likely responsible for most of the long wavelength 
emission. Higher resolution ISO observations for about a dozen mixed 
pairs revealed that the above working hypothesis was violated in many 
of the pairs. ISO maps, H$\alpha$ images and literature sources 
revealed pairs where AGN, LINER and starburst nuclei were present in 
the early-type component. In fact more Seyfert nuclei were found in the 
early-type components of our ISO mixed pairs sample (Domingue et al. 
2003) than in our ISO sample of n=8 S+S pairs (Xu et al. 2000). The ISO 
E+S and S+S samples were similarly selected for study on the basis of 
FIR brightness and signs of interaction. Since spiral galaxies are 
thought to preferentially harbor Seyfert galaxies we expected equal or 
greater numbers of active nuclei in our ISO S+S sample (16 spirals vs. 
12 spirals in the S+S and E+S samples, respectively). All galaxies in 
the S+S sample have good spectroscopic data available while a number of 
the E+S pairs have only a pre-CCD redshift determination. While the 
numbers are small, the ISO data raised the possibility of a significant 
population of active early-type galaxies in mixed pairs.

It is no longer surprising that interacting galaxies show evidence for 
enhanced non-thermal radio continuum emission because a surprisingly 
strong correlation exists between IR and radio flux measures in star 
forming galaxies (de Jong et al. 1985; Helou, Soifer, \& Rowan-Robinson 
1985). The correlation apparently reflects a close coupling between 
thermal and non-thermal processes related to star formation. Many radio 
continuum surveys of early-type galaxies now exist (Sadler, Jenkins, \& 
Kotanyi 1989; Wrobel \& Heeschen 1991; Slee et al. 1994) and, like IR 
results, (e.g. Knapp, Gunn, \& Wynn-Williams 1992) show low 
detection/flux levels. AGN often display excess radio emission which 
displaces them above the FIR-radio correlation (Condon \& Broderick 
1991). This makes the radio-FIR correlation a useful diagnostic tool 
for identifying AGN candidates in the absence of good quality nuclear 
spectra and high resolution IR data. Among early-type galaxies a 
tendency has been shown for S0's to obey the star formation relation 
(Bally \& Thronson 1989; Walsh et al. 1989; Wrobel \& Heeschen 1991) 
while ellipticals are less well defined and can show AGN 
characteristics (Wrobel \& Heeschen 1991). Our previous ISO results 
suggest that mixed pairs may represent an environment where active 
nuclei can be more easily generated via cross fueling of gas from the S 
component onto the slow rotating E component.

The results from the ISO studies motivated us to search for radio 
activity in a large sample of nearby mixed pairs by utilizing the NRAO 
VLA Sky Survey (NVSS, Condon et al.1998). The NVSS survey provides 
complete coverage for the Karachentsev (1972) isolated pair (KPG) and 
Karachentseva (1973) isolated single galaxy (KIG) catalogs. NVSS 
resolves the radio emission in virtually all of the pairs while the 
matching IRAS survey does not resolve the IR emission. Both surveys provide a reasonably large 
fraction of detected galaxies enabling us to construct useful radio-FIR 
correlation diagrams. This makes it possible for us to identify pair 
members that deviate from the star formation driven correlation. 

\section{Sample Selection and Analysis Procedure} 
The KPG is likely the most complete local sample of isolated binary 
galaxies and is estimated to be reasonably complete to 10$^{4}$ km 
s$^{-1}$ and m$_{Zw}$=15.0 (Xu \& Sulentic 1991). The catalog contains 
about 550 accordant redshift binaries with mixed pairs (E/S0+S) 
accounting for approximately 25-30\% of the sample. We restrict our 
study to mixed pairs with radial velocity $\leq$ 10$^{4}$ km s$^{-1}$ 
and component velocity difference $\Delta$v $\leq$ 10$^{3}$ km 
s$^{-1}$. The final adopted sample includes 112 mixed pairs. The 
optical positions of KPG components have a positional uncertainty 
between 1--10$\arcsec$ depending on the degree of central concentration 
in the galaxy.
The NVSS catalog was searched for all 5$\sigma$ source detections 
within 30$\arcsec$ of each galaxy in a mixed pair and radio-optical 
overlays assisted with identification. The NVSS survey is complete to 
S$_{1.4GHz}$=2.5 mJy and the rms positional uncertainty ranges from 
less than 1$\arcsec$ for S$_{1.4GHz}$$>$15 mJy up to 7$\arcsec$ for 
S$_{1.4GHz}$$<$15 mJy. The angular resolution of NVSS is $\sim$45$\arcsec$
at FWHM.  NVSS radio detections could immediately be 
assigned to one or both components in all but 20 of the pairs. 
Seventeen of the latter pairs with separation D $\leq$1 arcmin showed a 
radio position that was $\leq$0.2D closer to one component than the 
other making these assignments reasonably secure in almost all 
sources. KPG93 was confused because NED has reversed the 
identifications of the early and late-type components (data for the 
spiral component belongs to the early type and vice versa). The NVSS 
radio detection belongs to the Seyfert spiral component. The NVSS radio 
position in KPG116 lies slightly closer to the spiral and very close to 
a bright stellar object--it may be radio-loud quasar (it is listed as a 
spiral detection here). In many cases radio detection assignments 
could be confirmed with higher resolution FIRST (Becker, White, \& 
Helfand 1995) survey data (positional accuracy at or below 1$\arcsec$ at 
the survey threshold with FWHM of 5$\arcsec$). Sixteen of our pairs were included in FIRST and 
all of our NVSS detection assignments were confirmed. FIRST resolved 
two of the most ambiguous pairs involving KPG155 and 167. FIRST maps 
reveal that both components of KPG 155 are weak radio emitters but the 
early type is slightly stronger and the NVSS sources is centered on it. 
It was therefore assigned the NVSS detection. All of the emission from 
KPG167 originates from the compact early-type component (NED has them 
reversed). Table 1 summarizes the radio detection statistics and is 
formatted as follows: Column 1) KPG designation, 2) Pair Separation in arcmin
3) pair components detected, 4) NVSS flux for S detections, 5) NVSS flux for early-type detections, 6) adopted IRAS 60$\mu$m flux in Jy (Moshir et al. 1992) and 7) location of known spectroscopic ``activity'' 
listed as either the early-type (E) or the late-type (S) or both.

\section{1.4 GHz Detection Statistics} 
We define AGN here to include LINERs, Seyferts, Radio galaxies. We 
find that fourteen percent (detection fraction DF =0.14) of the pairs 
show radio emission from both components while the detection fractions 
for spirals and E/S0 components are DF=0.57 and DF=0.29, respectively. 
Radio continuum emission from 29\% of the early-types in pairs would 
have been a surprising result before the detection of MIR emission from 
many of them during the ISO survey (Domingue et al. 2003). Perhaps more 
surprising is that 24 of the 32 detected early-type components show a 
higher radio flux than their spiral companion.
If the early type components in mixed pairs are the same as isolated 
galaxies of similar types then we expect similar radio detection 
fractions. We adopt the Catalog of Isolated Galaxies (Karachentseva 
1973: KIG) as a control sample because it was compiled visually using 
an isolation criterion. In this sense it is a companion sample to the 
KPG and both samples show similar low galaxian density environments 
(typically, but not always, a nearest neighbor crossing-time of 
T$_c$$\sim$ 1 Gyr). If the morphologies can be trusted then the two 
samples differ only in the sense that the KPG early-types are involved 
in interacting binaries with a spiral companion. We recently 
reclassified all galaxies in the KIG using POSS2 and identify a sample 
of n=134 E/S0 galaxies of which x=25 are detected by NVSS yielding 
DF=0.18. A recent independent compilation of n=76 early-types from the 
KIG (Stocke et al. 2004) yields x=11 detections for a DF=0.14 
motivating us to adopt DF=0.16 for the control sample since both 
(overlapping) surveys have advantages and disadvantages. Adopting 
DF=0.16 as the probability for NVSS radio detection yields an expected 
sample of x=18 early-type KPG radio detections compared to X=32 
observed. The binomial probabilities for detecting 18 (expected) and 32 
(observed) early types out of a sample of 112 mixed pairs are 0.10 and 
3$\times$10$^{-4}$ respectively. Thus there is reasonable evidence for 
an excess of radio emitting early type KPG members.

Figure 1 shows the distribution of NVSS radio luminosities (H$_{0}$=75 km s$^{-1}$ Mpc$^{-1}$)
 for our KIG 
and KPG samples where we obtain mean values of log L$_{nvss}$= 21.5 
(n=25) and 21.9 (n=32) respectively. The distributions are not very 
Gaussian suggesting that a nonparametric rank-sum test is the more 
appropriate vehicle for comparing them. We obtain a rank-sum R$\sim$170 
and standard deviation s$\sim$60 giving us an approximately 3$\sigma$ 
difference in the sense that early-type members of KPG pairs are more 
radio luminous. We find almost twice the rate of radio detections and 
a factor $\times$3 higher radio luminosity for KPG early-type members 
compared to our KIG control sample. This result raises two questions: 
1) where did the gas giving rise to this effect come from and 2) is the 
excess radio emission a manifestation of star formation or AGN?

\section{FIR-Radio Continuum Correlations} 
The substantial number of both radio and FIR detections may provide an 
answer to the second of two questions posed above. We can resolve the 
radio emission in essentially all of the pairs while very few are 
resolved by IRAS. We expect that most of the FIR emission is due to 
star formation in the spiral components. We can make a first attempt at 
calibrating the star formation contribution in the pairs by deriving a 
radio-FIR correlation relation for pairs where only the spiral 
component shows an NVSS radio detection. It is safe to assume that any 
FIR contribution from an early-type galaxy is small if it is not 
detected by NVSS. The converse is not true. Figure 2a shows the 
radio-FIR relation for spiral-only pair NVSS detections. The strong 
correlation (de Jong et al. 1985; Helou et al. 1985; Reddy \& Yun 
2004) is thought to be driven by the relationship between thermal 
emission from dust heated by massive young stars and non-thermal 
emission related to the supernova demise of the same stars. Star 
formation is thought to be enhanced in KPG spirals because the FIR 
properties of both S+S and E/S0+S pairs are enhanced (Xu \&Sulentic 
1991; Toledo et al. 1999). Note that enhanced star formation will 
extend the correlation but little or no blurring is expected. The best 
fit regression line shown in Figure 2a gives a slope $\alpha$=0.84 with 
a CC=0.82 with $\sigma$=0.22. Four outlier spiral-only detections were 
not included in the fit shown in Figure 2a. They show excess radio 
emission (we know that the radio emission does not come from the 
early-type). Three of these galaxies (KPG 83, 419 and 526) show nuclear 
AGN/LINER spectra while KPG 116 involves the aforementioned radio 
quasar candidate. Our derived slope for the correlation is very similar 
to other estimates (e.g. $\alpha$=0.76 in Reddy \& Yun 2004) and is consistent with
the plotted star formation band in Fig.2 defined within the 3$\sigma$ deviations of the
flux-flux relation for the 2Jy IRAS sample of Yun, Reddy, \& Condon (2001).
We next consider pairs with an early-type radio detection. Figure 2b 
presents the radio-FIR correlation diagram for early-type detections 
in both the KIG (open boxes) and KPG (closed boxes) samples. Pairs 
included in Figure 2a are therefore not shown here. Points for the KPG 
sample indicate NVSS flux for the early type galaxy only. The 
vertical line indicates the contribution of the spiral component in 
pairs where both galaxies were detected. It is interesting that the KIG sample 
follows the star-formation band rather well with a slope for that 
sample of $\alpha$=0.64, CC=0.71 and standard deviation= 0.26. The 
simplest interpretation of the KIG correlation is that the S0 part of 
the sample shows low level star formation as suggested for other 
samples (e.g. Wrobel \& Heeschen 1991). The origin of the emission from 
elliptical galaxies is less clear. The KIG sample is taken as 
indicative of the typical radio/FIR properties for field early-type 
galaxies. Figure 2b shows the population of KPG early-type galaxies 
with higher than average radio fluxes. Only a few of the pairs where 
both components are NVSS detected (KPG 86, 127, 234, 476) show radio 
emission dominated by the spiral component. These pairs involve KPG 
spirals with very strong star formation. The residual radio emission 
from the early-type components in those pairs tends to be weaker than 
average. It is reasonable to assume that the spirals dominate the FIR 
flux even more and correction for it would move the early-type 
components towards the left and into the principal KIG occupation 
domain (log S$_{IRAS}$$<$3.0).

We adopt log S$_{nvss}$$>$1.0 as the zone of interest since only 5 KIG 
detections fall slightly above this level. KIG sources also follow the 
radio-FIR correlation more closely than the KPG early-types so we 
exclude sources with log S$_{iras}$$>$3.0 as likely to involve 
early-type components (S0) with FIR emission dominated by star 
formation. This zone boundary marks a conservative intersection of the outliers 
and the star formation band.

This leaves one KIG detection in the zone of interest 
compared to 11 early-type pair components (in eight cases only the 
early-type was detected by NVSS). An additional 6 of the latter class 
may move into the zone of interest after correction for IRAS flux from 
the spiral components (KPG sources with log S$_{IRAS}$$>$ 3.25 other 
than KPG 86, 127, 234 and 476). It is therefore possible that about 17 
early-type pair components show unusually strong radio emission 
relative to the KIG defined radio-FIR relation. Ten known AGN 
(radio galaxy/Seyfert/LINER) are already known among this early-type 
population with many of the others lacking any modern spectroscopic 
data. Note that the radio-FIR correlation will only help to find AGN 
that show above average radio emission. Some of the known AGN in our 
sample fall at or below the lower edge of the zone of interest. A 
spectroscopic survey, especially of the remaining candidate pairs in 
the zone of interest, will likely reveal additional active nuclei among 
the early-type detections. Comparison of NVSS and FIRST radio data 
supports this expectation. While NVSS resolution is too low to resolve 
the emission within a pair component, it is likely to detect all of the 
radio flux. FIRST, on the contrary, is insensitive to extended emission 
due to spatial frequency attenuation in that VLA survey. Typically we 
find a large difference between peak and integrated flux for spiral 
FIRST detections of spiral components indicating resolution of the 
source. The resolved emission is likely related to star formation 
leaving, in many cases, little unresolved flux that could be assigned 
to a nuclear AGN. Most of the FIRST detections of early-type components 
show near equality between NVSS, peak FIRST and integrated FIRST radio 
fluxes. This indicates that most of the radio emission is from 
early-types is unresolved which is consistent with an AGN-related radio 
signature.

The excess of early-type radio-detections in pairs motivates serious 
consideration of an interaction related explanation. Gas transfer from 
the gas-rich spiral to the early-type (i.e. cross-fueling) component 
has been discussed for many years (Haynes, Giovanelli, \& Chincarini 
1984; Demin 1984; Sotnikova 1988ab; Sulentic 1992; Anderson, Sulentic, 
\& Rampazzo 1994; de Mello et al. 1996; ). Some of the pairs have been 
discussed in this context including KPG468 one of the most FIR luminous 
pairs in our sample. In this case correction for radio/FIR emission 
from the spiral does not significantly change the position of the 
source in Figure 2b. The early-type companion involves: 1) both 
starburst and AGN signatures while 2) the underlying morphology suggests 
that it is an early-type and 3) the known kinematical data allow for the 
possibility of cross-fueling (Jenkins 1984; Gonzalez-Delgado et al. 
1996; Hernandez-Toledo et al. 2003). The activity in the early-type 
component of KPG234 is not yet extreme but shows evidence consistent 
with mass transfer (Gonzalez-Delgado et al. 1997). In the case of 
KPG591 kinematical data and modeling suggest that mass transfer has 
taken place (Marcelin et al. 1987; Salo \& Laurikainen 1993). In some 
of these cases the early-type companion is likely S0 rather than 
elliptical. Examples of cross fueling might be more difficult to 
identify in such cases because the KIG control sample suggests that 
they are more likely to show radio and FIR emission from low level star 
formation activity.

The most surprising result involves the number of pairs where ONLY the 
early-type component is detected in NVSS. At least nine of the pairs in 
our zone of interest fall in this class including two radio-galaxies 
(KPG303 and 445). How can one cross fuel if the donor spiral shows no 
significant ISM signature? In at least one case (KPG93) recent HST 
images appear to show {\it in flagrante delicto} cross fueling from, in 
this case, an active Seyfert spiral onto an S0 manifesting low level 
LINER activity (Keel 2004). In this case the pair separation challenges 
the resolving power of NVSS. However the accuracy of the NVSS radio 
position for a 15 mJy detection is very high and suggests that most of 
the radio emission originates in the early-type component (Condon et 
al. 2002 note that one cannot rule out a contribution from the spiral). 
The same can be said for very close pairs KPG 460 and 487. In the other 
cases the separation allows no ambiguity in the detection assignment. 
Is the spiral misclassified in these cases (e.g. an "early-type" pair; 
Rampazzo \& Sulentic 1992)? Our unpublished CCD images for all of these 
pairs show evidence for spiral structure and allow only KPG 383 to 
be questioned as a true mixed pair. The more likely explanation is that 
the spirals in early-type only NVSS detected pairs are experiencing low 
levels of star formation. Perhaps donor spirals become anemic by cross 
fueling their companion and nuclear fueling of their own AGN in 
response to tidal effects of the galaxy being fueled (e.g. KPG93)? 
Both radio galaxies in our sample (KPG303 and 445) show companions with 
a well defined disk but little evidence for HII regions or dust lanes. 
They are the most anemic "spirals'' in our sample. We are not 
suggesting that all of our early-type detections are examples of cross 
fueling. Some may simply involve slow rotators with a small amount of 
gas that is more rapidly channeled into their nuclei through the 
dynamical effect of the spiral.

\section{Summary and Discussion} 
We find a significant excess of mixed pairs with early-type NVSS 
detections that also show a $\times$3 enhancement in average radio 
luminosity. The detections involve both E and S0 companions of spiral 
galaxies. We suggest that this excess can be interpreted as a 
population of early-types cross fueled by their gas rich neighbors. A 
surprising number already show evidence of nuclear activity including 
AGN/LINER/starburst while modern spectroscopic data exists for only 
about half of them. We suggest that cross fueling onto slow rotators 
will more quickly fuel nuclear AGN activity than in spirals. We may be 
observing a sequence of activity beginning with cross fueling that 
leads to enhanced star formation activity flowed by ignition af an 
AGN. At least two of our early type detections involve hybrid 
starburst/AGN activity. The two most extreme cases involving radio 
galaxies show disky companions that might be interpreted as anemic 
spirals or lenticulars. These may represent the end cycle of the 
process where AGN activity leads to development of radio loudness in 
the last stages of the AGN process. KPG445 shows a double-jet radio 
morphology consistent with the early stages in the development of 
double-lobed (FRII) structure. 

This research has made use of the NASA/IPAC Extragalactic Database (NED), which 
is operated by JPL, Caltech, under contract with the NASA.
 
\clearpage
\begin{figure}
\figcaption{Histogram of radio luminosity for the 32 early-type KPG's detected by NVSS and for the 25 early-type KIG's NVSS detected.} 
\end{figure} 
\clearpage

\begin{figure}
\figcaption{(a) NVSS flux vs. 60$\mu$m IRAS flux in log-log scale for 
the 46 pairs with only a spiral detection. We use these pairs to 
determine a Radio/FIR correlation for star forming galaxies of this 
sample. The 4 labeled pairs are excluded from the fit. The solid lines mark the region of star forming galaxies defined from a 3sigma deviation to the correlation of Yun, Reddy, \& Condon 2001.
(b) The same as (a) for pairs with detected early-types (closed circles) and 
star formation band as in (a). Also plotted are the early-type galaxies(open 
squares) from the KIG control sample. Dashed vertical lines display 
the pairs before and after removing the spiral contribution from the 
total NVSS flux.}
\end{figure}

\clearpage
\oddsidemargin=-1cm
\tabletypesize{\scriptsize}

\begin{deluxetable}{lrcrrrc} 
\tablecolumns{5} 
\tablewidth{0pt} 
\tablecaption{Mixed Pairs with Radio Detections.} 
\tablehead{ \colhead{KPG}&\colhead{Sep}&\colhead{NVSS}&\colhead{S$_{1.4 GHz}$(S)}&\colhead{S$_{1.4 GHz}$(E/S0)}&IRAS S$_{60}$&\colhead{AGN}\\ 
\colhead{Name}&\colhead{($\arcmin$)}&\colhead{Detection}&\colhead{(mJy)}&\colhead{(mJy)}&\colhead{(Jy)}&\colhead{Location} }
\startdata

078 &3.50 & both& 64.7 &15.9& 3.55&\ldots\\ 
086&3.48 & both& 26.4& 4.6& 4.32 &E\tablenotemark{1}\\ 
127&6.32 & both &267.0& 2.9& 12.00 &\ldots\\ 
188&1.23  &both &3.0& 3.9& 1.09 &\ldots\\ 
202&1.29  &both &24.7& 7.0& 2.27 &\ldots\\ 
229&4.57  &both &6.7& 31.1& 0.59 &\ldots\\ 
234&2.35  &both &97.5& 3.2& 7.83 &E\&S\tablenotemark{1}\\ 
353&2.57  &both &54.4& 29.1& 6.11 &\ldots\\ 
436&1.21  &both &7.4& 4.0& 0.83 &\ldots\\ 
468& 0.77 &both &6.3& 91.4& 10.00 &E\&S\tablenotemark{2}\\ 
476&1.98  &both &11.9& 19.2& 2.70 &\ldots\\ 
508&3.79  &both &4.7& 145.1& 0.34 &E\tablenotemark{3}\\ 
530&9.27 &both &4.4& 14.6& 0.40 &\ldots\\ 
552&1.23  &both &5.5& 16.9& 2.80 &E\tablenotemark{4}\\ 
553&1.42  &both &5.8& 6.3& 0.35 &\ldots\\ 
591&2.00 &both &16.9& 24.9& 4.80 &\ldots\\ 
062&2.39  &E& \ldots& 11.9& 0.22 &\ldots\\ 
093&0.21  &E& \ldots&15.2& 0.87 &E\&S\tablenotemark{5}\\ 
130&0.53  &E& \ldots& 8.1& 1.34 &\ldots\\ 
155&0.33  &E& \ldots& 3.8& 0.52 &\ldots\\ 
167&0.44  &E& \ldots& 3.1& 0.48 &\ldots\\ 
239&3.92  &E& \ldots& 10.7& 0.18 &\ldots\\ 
303&2.03  &E& \ldots& 481.4& 0.14 &E\tablenotemark{6}\\ 
383&0.45  &E& \ldots& 39.2& 0.50 &\ldots\\ 
386&1.97  &E& \ldots&31.5& 0.50 &\ldots\\ 
445&0.87  &E& \ldots&385.3& 0.14 &E\tablenotemark{7}\\ 
460&0.79  &E& \ldots&15.3& 0.13 &\ldots\\ 
485&0.35  &E& \ldots& 3.5& 0.20 &\ldots\\ 
487&0.32  &E& \ldots& 12.9& 1.85 &\ldots\\ 
509&2.15  &E& \ldots& 11.5& 0.33 &\ldots\\ 
536&0.42  &E& \ldots& 20.2& 1.49 &E\tablenotemark{4}\\ 
537&1.37  &E& \ldots& 2.5& 0.24 &\ldots\\ 
061& 2.02 &S& 10.0&\ldots& 0.54 &\ldots\\ 
074& 0.68&S& 5.6 &\ldots& 0.44 &\ldots\\ 
079&1.02  &S& 3.8 &\ldots& 1.11 &\ldots\\ 
081&4.58  &S& 7.7 &\ldots& 0.42 &\ldots\\ 
083&0.71  &S& 154.6 &\ldots& 5.06 &S\tablenotemark{4}\\ 
089&1.51  &S& 2.4 &\ldots& 0.38 &\ldots\\ 
101&0.67  &S& 4.4 &\ldots& 0.49 &\ldots\\ 
116&1.57  &S& 213.1 &\ldots& 0.84 &\ldots\\ 
121&2.08  &S& 8.2 &\ldots& 0.57 &\ldots\\ 
122&1.34  &S& 7.2 &\ldots& 0.66 &\ldots\\ 
128&3.30  &S& 23.2 &\ldots& 1.99 &\ldots\\ 
134&0.60  &S& 7.4 &\ldots& 0.41 &\ldots\\ 
144&0.54  &S& 10.2 &\ldots& 1.38 &\ldots\\ 
153&0.54  &S& 2.8 &\ldots& 0.40 &\ldots\\ 
182&0.92  &S& 4.8 &\ldots& 0.79 &\ldots\\ 
198&1.31  &S& 8.4 &\ldots& 0.66 &\ldots\\ 
209&0.45  &S& 22.7 &\ldots& 0.68 &\ldots\\ 
243&1.08  &S& 9.1 &\ldots& 0.93 &\ldots\\ 
248&0.77  &S& 19.8 &\ldots& 1.52 &\ldots\\ 
269& 1.07 &S& 21.5 &\ldots& 1.37 &\ldots\\ 
275& 1.57 &S& 17.0 &\ldots& 0.50 &\ldots\\ 
284&2.68  &S& 15.3 &\ldots& 1.02 &\ldots\\ 
304&1.68  &S& 9.4 &\ldots& 0.69 &E\tablenotemark{8}\\ 
317& 2.25 &S& 2.8 &\ldots& 0.28 &\ldots\\ 
363& 0.37 &S& 38.8 &\ldots& 5.73 &S\tablenotemark{4}\\ 
380&0.91  &S& 4.0 &\ldots& 0.19 &\ldots\\ 
393&2.38  &S& 13.6 &\ldots& 1.03 &\ldots\\ 
394&1.57  &S& 23 &\ldots& 1.81 &\ldots\\ 
407&0.32  &S& 5.8 &\ldots& 0.76 &\ldots\\ 
412&6.15  &S& 2.2 &\ldots& 0.28 &\ldots\\ 
416&3.21  &S& 29.0 &\ldots& 3.36 &\ldots\\ 
419&3.82  &S& 338.8 &\ldots& 8.41 &S\tablenotemark{4}\\ 
429&4.31  &S& 4.8 &\ldots& 0.61 &\ldots\\ 
432&3.35  &S& 3.6 &\ldots& 0.70 &\ldots\\ 
439& 0.86 &S& 35.3 &\ldots& 3.69 &\ldots\\ 
451&5.61  &S& 29.4 &\ldots& 5.11 &\ldots\\ 
465&4.00  &S& 9.1 &\ldots& 1.08 &\ldots\\ 
519&3.03  &S& 2.9 &\ldots& 0.53 &\ldots\\ 
526& 2.31 &S& 182.2 &\ldots& 0.64 &S\tablenotemark{4}\\ 
542&0.68  &S& 29.2 &\ldots& 2.47 &\ldots\\ 
547&0.67  &S& 17.7 &\ldots& 1.73 &\ldots\\ 
570&5.20  &S& 45.7 &\ldots& 0.21 &\ldots\\ 
572&0.92  &S& 13.1 &\ldots& 0.50 &\ldots\\ 
576& 1.55 &S& 2.6.0 &\ldots& 0.38 &\ldots\\ 
583&1.00  &S& 3.0 &\ldots& 0.62 &\ldots\\ 
596&0.67  &S& 10.0 &\ldots& 1.82 &\ldots\\ 
\enddata 
\tablerefs{(1) Ho et al. 1995; (2) Gonz\'alez-Delgado \& P\'erez 1996; (3) Corbett et al. 1998; (4) Keel 1996; (5) Keel 2004; (6) Puschell 1981; (7) Liu \& Zhang 2002; (8) Magliocchetti et al. 2002} 
\end{deluxetable}

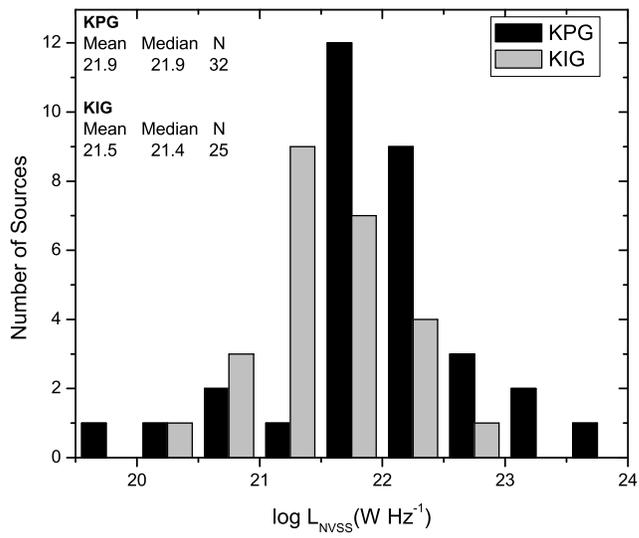
\end{document}